\documentclass{mpe_report}

\usepackage{psfig,graphicx,epsfig}
\usepackage{color}
\usepackage{amsmath,amssymb,epic,eepic,array}

\begin{document}

\pagenumbering{arabic}
\setcounter{page}{20}

\renewcommand{\FirstPageOfPaper }{ 20}\renewcommand{\LastPageOfPaper }{ 23}

\def\be{\begin{equation}}
\def\ee{\end{equation}}
\def\lesssim{\raisebox{-0.3ex}{\mbox{$\,\, \stackrel{<}{_\sim} \,\,$}}}
\def\gtrsim{\raisebox{-0.3ex}{\mbox{$\stackrel{>}{_\sim} \,$}}}
\def\EB{\hbox{${\rm {\bf E} \times {\bf B}}$}}

\title{Thermal radiation from hot polar cap in pulsars}
\author{Janusz Gil\inst{1,2}\and George Melikidze\inst{1,3} \and Bing Zhang\inst{2}}
\institute{Institute of Astronomy, University of Zielona G\'ora,
Lubuska 2, 65-265, Zielona G\'ora, Poland \and Department of
Physics,University of Nevada Las Vegas, Las Vegas, NV, USA \and
Abasumani Astrophysical Observatory, Al. Kazbegi ave. 2a, 0160,
Tbilisi, Georgia} 

\maketitle

\begin{abstract}
 Thermal radiation from hot polar caps is examined in radio pulsars
 with drifting subpulses. It is argued that if these subpulses
 correspond to sparking discharges of the inner acceleration region right
 above the polar cap surface then a
 simple relationship between the observed subpulse drift rate in radio
  and thermal X-ray luminosity from the polar cap heated
 by sparks should exist. This relationship is derived and tested
 in pulsars for which an appropriate good quality data is
 available.
\end{abstract}


\section{Introduction}

Almost 40 years have passed since the discovery of pulsars and the
mechanism of their coherent radio emission is still not known. The
puzzling phenomenon of drifting subpulses is widely regarded as a
powerful tool for the investigation of mechanisms of pulsar radio
emission. In the classical model of Ruderman \& Sutherland (1975;
RS75 henceforth) the spark-associated subbeams of subpulse
emission circulate around the magnetic axis due to
$\mathbf{E}\times\mathbf{B}$ drift of spark plasma filaments. This
model is widely regarded as the most natural and plausible
explanation of drifting subpulse phenomenon, at least
qualitatively. Despite its popularity the RS75 model is known to
suffer from the so-called binding energy problem (for review see
Usov \& Melrose 1995, 1996). In fact, the cohesive energy of
surface iron ions were largely overestimated in RS75 and the
"vacuum gap" envisioned by RS75 was impossible to form. A number
of attempts have been made to resolve this problem, including a
quite exotic proposal that radio pulsars with drifting subpulses
are bare polar cap strange stars (BPCSS) rather than neutron stars
(Xu, Qiao \& Zhang, 1999). Gil \& Melikidze, (2002) argued that
the formation of RS75 gap above the neutron star polar cap was, in
principle, possible, although it required a very strong surface
magnetic field, much stronger than the dipolar component inferred
from the observed spin-down rate. Growing evidence of such strong
non-dipolar surface field accumulates in the literature, both
observational and theoretical (see Urpin \& Gil 2004, for short
review).

Even if the RS75 gap was possible to form, it would not
automatically solve the mystery of drifting subpulses. It is well
known that the original model of RS75 predicts too fast a drifting
rate (e.g. Deshpande \& Rankin, 1999; DR99 henceforth). Motivated
by this issue Gil, Melikidze \& Geppert (2003; GMG03 henceforth)
developed further the idea of the inner acceleration region above
the polar cap by including the partial screening due to thermionic
ions flow from the surface heated by sparks. We will call this
kind of the inner acceleration region the ``partially screened gap"
(PSG henceforth). The slow drift corresponds to \EB\ drift of
spark generated electron-positron pairs, until the total charge
density within the partially screened gap reaches the
co-rotational value. Since the PSG potential drop is much lower
than in the RS75 model, the intrinsic drift rate is compatible
with observations (for details see GMG03).

A distinguishing property of PSG model is relatively high
predicted heating rate of the polar cap surface, compatible with
observations (in contrast to RS75 gap, which overheated the polar
cap). The space charge limited model (Arons \& Scharleman, 1979,
AS79 henceforth; Zhang \& Harding, 2000; Harding \& Muslimov,
2002, HM02 henceforth) predicts a much lower polar cap heating
rate. On the other hand, in the BPCSS model (Xu, Qiao, \& Zhang,
1999) no hot spot is expected due to the high thermal conductivity
at the BSS surface. Thus, measuring the thermal X-ray luminosity
from heated polar caps can potentially reveal the nature of the
inner acceleration region in pulsars. This can help us to
understand a mechanism of drifting subpulses, which appears to be
a common phenomenon in radio pulsars. Recently Weltevrede,
Edwards, \& Stappers (2006, WES06 henceforth) presented results of
a systematic, unbiased search and found that the fraction of
pulsars showing drifting subpulses is at least 55~\%. They
concluded that the conditions for drifting mechanism to work
cannot be very different from the emission mechanism of radio
pulsars.

In order to test different available models of inner acceleration
region in pulsars Zhang, Sanwal \& Pavlov 2005; ZSP05 henceforth)
observed the best studied drifting subpulse radio pulsar PSR
B0943$+$10 with the {\em XMM-Newton} observatory. Their
observations were consistent with PSG formed in strong,
non-dipolar magnetic field just above the surface of very small
polar cap. Recently, Kargaltsev, Pavlov \& Garmire (2006, KPG06
henceforth) observed the X-ray emission from the nearby PSR
B1133$+$16 and found that this case is also consistent with the
thermal radiation from a small hot spot (again much smaller than
the canonical polar cap). PSR B1133$+$16 is almost a twin of PSR
B0943$+$10 in terms of $P$ and $\dot{P}$ values and,
interestingly, both these pulsars have very similar X-ray
signatures, in agreement with our PSG model (see Table 1).

In this paper we generalized the treatment of ZSP05 and developed
detailed model for thermal X-ray emission from radio drifting
pulsars. The model matches the observations of PSRs B0943+10 and
B1133+16 well, both in radio and X-rays. There is a number of
pulsars with measured thermal X-ray radiation from small hot polar
cap but unfortunately their drifting properties are not yet known.
This is likely to change in the near future due to new
sophisticated methods for analysis of intensity fluctuations in
weak pulsars being developed (e.g. WRS06 and references therein).

\section{PSG model of the inner acceleration region}

As already mentioned, growing evidence appears, both
observationally and theoretically, that the actual surface
magnetic field $B_s$ is highly non-dipolar. Its magnitude can be
described in the form $B_s=bB_d$ (Gil \& Sendyk 2000; GS00
henceforth), where the enhancement coefficient $b>1$ and
$B_d=2\times 10^{12}(P\dot{P}_{-15})^{1/2} {\rm G}$ is the
canonical, star centered dipolar magnetic field, $P$ is the pulsar
period and $\dot{P}_{-15}=\dot{P}/10^{-15}$ is the period
derivative.

The polar cap is defined as the locus of magnetic field lines that
penetrate the so-called light cylinder. Conventionally, the polar
cap radius $r_{pc}=1.45\times 10^4P^{-0.5}~{\rm cm}$, and its
surface area $A_{pc} \sim 2\times 10^8 P^{-1}$ cm$^2$. In the case
of non-dipolar surface field the polar cap area must shrink due to
flux conservation of the open field lines $A_{pc} B_d=A_p B_s$.
Thus, the surface area of the actual polar cap $A_p=b^{-1}
A_{pc}$, regardless of the actual shape of the polar cap.
Consequently, one can write the polar cap radius in the form
$r_p=b^{-0.5}r_{pc}$ (GS00), realizing however that this is only a
characteristic dimension of the actual polar cap. First, the
canonical expression for the radius of the polar cap (GJ69) is
only a geometrical approximation (e.g. Michel 1973). Secondly, the
presence of strong surface magnetic field anomalies should make
the shape of the polar cap quite irregular.

The charge depleted inner acceleration region above the polar cap
results from the deviation of a local charge density $\rho$ from
the co-rotational charge density (Goldreich \& Julian 1969)
$\rho_{\rm GJ}=-{\mathbf\Omega}\cdot{\bf B}_s/{2\pi
c}\approx{B_s}/{cP}$. For isolated neutron stars one might expect
the surface to consist mainly of iron formed at the neutron star's
birth (e.g. Lai 2001). Therefore, the charge depletion above the
polar cap can result from bounding of the positive $^{56}_{26}$Fe
ions (at least partially) in the neutron star surface. As
demonstrated by recent exact calculations of Medin \& Lai (2006),
the iron chains are strongly bound in magnetic field close to
$10^{14}$ G. This can only occur if the surface magnetic field in
actual pulsars is dominated by strong non-dipolar components. If
this is really possible, then due to significant bounding the
positive charges cannot be supplied at the rate that would
compensate the inertial outflow through the light cylinder. As a
result, a significant part of the unipolar potential drop develops
above the polar cap, which can accelerate charged particles to
relativistic energies and power the pulsar radiation mechanism.

The ignition of cascading production of electron-positron plasma
is crucial for limitation of growing gap potential drop. The
accelerated positrons will leave the acceleration region, while
the electrons will bombard the polar cap surface, causing a
thermal ejection of ions. This thermal ejection will cause partial
screening of the acceleration potential drop $\Delta V$
corresponding to a shielding factor $\eta=1-\rho_{i}/\rho_{\rm
GJ}$ (see GMG03 for details), where $\rho_{i}$ is charge density
of ejected ions, $\Delta V=\eta({2\pi}/{cP})B_s h^2$ is the
potential drop and $h$ is the height of the acceleration region.
The gap potential drop is completely screened when the total
charge density $\rho=\rho_i+\rho_+$ reaches the co-rotational
value $\rho_{GJ}$.

GMG03 argued that the actual potential drop $\Delta V$ should be
thermostatically regulated and the quasi-equilibrium state should
be established, in which heating due to electron bombardment is
balanced by cooling due to thermal radiation. The
quasi-equilibrium condition is $Q_{cool}=Q_{heat}$, where
$Q_{cool}=\sigma T_s^4$ is a cooling power surface density by
thermal radiation from the polar cap surface and $Q_{heat}=\gamma
m_ec^3n$ is heating power surface density due to back-flow
bombardment, $\gamma=e\Delta V/m_ec^2$ is the Lorentz factor,
$n=n_{GJ}-n_{i}=\eta n_{GJ}$ is the number density of back-flowing
plasma particles depositing their kinetic energy at the polar cap
surface, $\eta$ is the shielding factor, $n_{i}$ is the charge
number density of thermionic ions and
$n_{GJ}=\rho_{GJ}/e=1.4\times
10^{11}b\dot{P}_{-15}^{0.5}P^{-0.5}{\rm cm}^{-3}$ is the
corotational charge number density. It is straightforward to
obtain an expression for the quasi-equilibrium surface temperature
in the form
\be
T_s=(6.2\times 10^4{\rm K})(\dot{P}_{-15}/{P})^{1/4}\eta^{1/2}b^{1/2}h^{1/2}.
\label{Ts}
\ee

Let us now interrelate the accelerating potential drop $\Delta V$
and the perpendicular (with respect of the magnetic field lines)
electric field $\Delta E$ which causes \EB\ drift. Following the
original method of RS75 we can argue that the tangent electric
field is strong only at the polar cap boundary where $\Delta
E=0.5{\Delta V}/{h}=\eta({\pi}/{cP})B_sh$ (see Appendix~A in GMG03
for details). Due to the \EB\ drift the discharge plasma performs
a slow circumferential motion with velocity $v_d=c\Delta
E/B_s=\eta\pi h/P$. The time interval to make one full revolution
around the polar cap boundary is $\hat{P}_3\approx 2\pi r_p/v_d$.
One then has
\be
\frac{\hat{P}_3}{P}=\frac{r_p}{2\eta h}.
\label{P3P}
\ee
If the plasma above the polar cap is fragmented into filaments
(sparks) which determine the intensity structure of the
instantaneous pulsar radio beam, then in principle, the tertiary
periodicity $\hat{P}_3$ can be measured/estimated from the pattern
of the observed drifting subpulses (Deshpande \& Rankin 1999, Gil
\& Sendyk 2003). According to RS75, $\hat{P}_3=NP_3$, where $N$ is
the number of sparks contributing to the drifting subpulse
phenomenon observed in a given pulsar and $P_3$ is the primary
drift periodicity (distance between the observed subpulse drift
bands). On the other hand $N\approx 2\pi r_p/2h$ (GS00). Thus, one
can write the shielding factor in the form $\eta\approx
(1/2\pi)(P/P_3)$ , which depends only on an easy-to-measure
primary drift periodicity. Apparently, the shielding parameter
$\eta$ should be much smaller than unity.

\begin{table*}
\begin{minipage}{170mm}
\begin{center}

\fontsize{10}{10pt}\selectfont

\caption{Comparison of observed and predicted parameters of
thermal emission from hot polar caps}
\begin{tabular}{c c c c c c c c c c c}
\hline \hline Name & $P_3/P$ & \multicolumn{2}{c}{$\hat{P}_{3}/P$}
& \multicolumn{2}{c}{${L_{\mathrm{x}}}/{\dot{E}}\times 10^{3}$} &
$b$ & $T_{s}^{\mathrm{(obs)}}$ &
$T_{s}^{\mathrm{(pred)} }$ & $B_{\mathrm{d}}$ & $B_{\mathrm{s}}$ \\
\hline \ PSR B & Obs. & Obs. & Pred. & Obs. & Pred. &
${A_{\mathrm{pc}}}/{A_{\mathrm{bol}}}$ & $10^{6}\ $K & $10^{6}\ $K & $10^{12}$G & $10^{14}$G \\
\hline $0943+10$& 1.86 & $37.4$ & $36_{-2}^{+8}$ &
$0.49_{-0.16}^{+0.06}$ & $0.45$ & $60_{-48}^{+140}$ &
$3.1_{-1.1}^{+0.9}$ & $3.3_{-1.1}^{+1.2}$ & $3.95$ & $
2.37_{-1.90}^{+5.53}$
\\
$1133+16$ & $3^{+2}_{-2}$ & ($33_{-3}^{+3}$) & $27_{-2}^{+5}$ &
$0.77_{-0.15}^{+0.13}$ & $0.58_{-0.09}^{+0.12}$  &
$11.1_{-5.6}^{+16.6}$ & $2.8_{-1.2}^{+1.2}$ & $2.1_{-0.4}^{+0.5}$
&
$4.25$ & $ 0.47_{-0.24}^{+0.71}$ \\
\hline
\end{tabular}
\end{center}
\end{minipage}
\end{table*}

The X-ray thermal luminosity is $L_x=\sigma T_s^4\pi
r_p^2=1.2\times 10^{32}(\dot{P}_{-15}/P^3)(\eta h/r_p)^2$~erg/s,
which can be compared with the spin-down power
$\dot{E}=I\Omega\dot{\Omega}=3.95 I_{45}\times
10^{31}\dot{P}_{-15}/P^3$~erg/s, where $I=I_{45}10^{45}$g\ cm$^2$
is the neutron star moment of inertia (bellow we assume that
$I_{45}=1$). Using equation~(\ref{P3P}) we can derive the thermal
X-ray luminosity and its efficiency as $L_x=2.5\times
10^{31}(\dot{P}_{-15}/P^3)(P/\hat{P}_3)^{2}$, or in the simpler
form representing the efficiency with respect to the spin-down
power
\be \frac{L_x}{\dot{E}}=0.63 \left(\frac{P}{\hat{P}_3}\right)^{2}
\label{Lx},
\ee which is very useful for comparison with observations. One
should realize that this equation holds only for thermal X-rays
from hot spot and cannot be applied neither to cooler radiation
from the entire stellar surface nor to the magnetospheric
component.

We can see that $L_x$ in these equations depends only on radio
observables. It is particularly interesting and important that
both equations above do not depend on details of the sparking gap
model $(\eta, b, h)$. Although one has to be careful whether all
our assumptions are satisfied in real pulsars, it seems that we
have found a very useful, relatively easy testable (at least for
the order-of-magnitude) relationship between the properties of
drifting subpulses observed in radio band and the characteristics
of thermal X-ray emission from the polar cap heated by sparks
associated with these subpulses. For PSRs B0943$+$10 and B1133+16,
which are the only two pulsars in which both $\hat{P}_3$ and $L_x$
are measured/estimated (Table~1), the above equation holds very
well. Interestingly, in both cases $L_x/\dot{E} \sim 10^{-3}$.

Using equation~(\ref{P3P}) we can write the polar cap temperature in the form
\be
T_s=(5.1\times 10^6 {\rm
K})b^{1/4}\dot{P}^{1/4}_{-15}P^{-1/2}\left(\frac{\hat{P}_3}{P}\right)^{-1/2}
\label{Ts4},
\ee
where the enhancement coefficient $b=B_s/B_d\approx
A_{pc}/A_{bol}$, $A_{pc}=\pi r^2_{pc}$ and $A_{bol}=A_p$ is the
actual emitting surface area (bolometric). Since $A_{bol}$ can be
determined from the black-body fit to the spectrum of the observed
hot-spot thermal X-ray emission, the above equations can be
regarded as independent of details of the sparking gap model and
depending only on combined radio and X-ray data, similarly as in
equations (\ref{Lx}).

\section{Comparison with observational data}

Table 1 presents the observational data and predicted values of a
number of quantities for two pulsars, which we believe show clear
evidence of thermal X-ray emission from the spark heated polar
caps as well as they have known values of tertiary subpulse drift
periodicity. The predicted value of $\hat{P}_3$ and/or $L_x$ were
computed from eq.(4), while the predicted values of $T_s$ were
computed from eq.(5), with $b=A_{pc}/A_{bol}$ determined
observationally. The actual surface magnetic field $B_s=bB_d$,
where $B_d$ is the canonical dipole magnetic field at the polar
cap. As a principle, the predicted (Pred) values were obtained
from observational (Obs) values through Eq.(3).

{\it\bf PSR B0943$+$10.} This is the best studied drifting
subpulse radio pulsars with $P=1.09$~s, $\dot{P}_{-15}=3.52$,
$\dot{E}=10^{32}\ {\rm erg\ s}^{-1}$, $P_3=1.86P$,
$\hat{P}_3=37.4P$ and $N=\hat{P_3}/P_3=20$ (DR99). It was observed
by ZSP05, who obtained an acceptable thermal BB fit with
bolometric luminosity $L_x=(5^{+0.6}_{-1.6})\times 10^{28}\ {\rm
erg\ s}^{-1}$ and thus $L_x/\dot{E}=(0.49^{+0.06}_{-0.16})\times
10^{-3}$. The bolometric polar cap surface area
$A_{bol}=10^7[T_s/(3\times 10^6 {\rm K})]^{-4} {\rm
cm}^2\sim(1^{+4.0}_{-0.4})\times 10^7\ {\rm cm}^2$ is much smaller
than the conventional polar cap area $A_{pc}=6\times 10^8\ {\rm
cm}^2$. This all correspond to the best fit temperature $T_s \sim
3.1\times 10^6$~K (see Fig.~1 in ZSP05). The predicted value of
$L_x/\dot{E}$ calculated from equation~(\ref{Lx}) agrees very well
with the observational data. The surface temperature $T_s$
calculated from equation~(\ref{Ts4}) with $b=A_{pc}/A_{bol}$ is
also in good agreement with the best fit. The shielding factor
$\eta=0.09$, thus more than 90 \% of the available vacuum
potential drop is screened.

{\it\bf\ PSR B1133$+$16.} This pulsar with $P=1.19$~s,
$\dot{P}_{-15}=3.7$, and $\dot{E}=9\times 10^{31}\ {\rm erg\
s}^{-1}$ is almost a twin of PSR B0943$+$10. KPG06 observed this
pulsar with Chandra and found an acceptable BB fit
$L_x/\dot{E}=(0.77^{+0.13}_{-0.15})\times 10^{-3}$,
$A_{bol}=(0.5^{+0.5}_{-0.3})\times 10^7~{\rm cm}^2$ and
$T_s\approx 2.8\times 10^6$~K. These values are also very close to
those of PSR B0943$+$10, as should be expected for twins. Using
equation (\ref{Lx}) we can predict $\hat{P}_3/P=27^{+5}_{-2}$ for
B1133+16. Interestingly, Nowakowski (1996) obtained fluctuation
spectrum for this pulsar with clearly detected long period feature
corresponding to about $32P$. Most recently, WES06 found
$P_3/P=3\pm 2$ and long period feature corresponding to $(33\pm 3)
P $ in the fluctuation spectrum of PSR B1133+16. The latter value
seem to coincide with that of Nowakowski (1996), as well as with
our predicted range of $\hat{P}_3$. We therefore claim that this
is the actual tertiary periodicity in PSR B1133+16 and show it in
paranthesis in Table 1.
It is worth noting that $\hat{P}_3/P_3=33\pm 3$ is quite close to
37.4 measured in the radio twin PSR B0943+10. Note also that the
number of sparks predicted from our hypothesis is
$N=\hat{P}_3/P_3=(33\pm 3)/(3\pm 2)=11^{+25}_{-6}$, so it can also
be close to 20, as in the case of twin PSR B0943$+$10. The
shielding factor $\eta=0.05^{+0.11}_{-0.02}$. Thus, only several
\% of the vacuum gap potential drop is available for acceleration.

\section{Conclusions and discussion}

Within the partially screened gap model of the inner acceleration
region in pulsars developed by GMG03 we derived a simple
relationship between the X-ray luminosity $L_x$ from the polar cap
heated by sparks and the tertiary periodicity $\hat{P}_3$  of the
spark-associated subpulse drift observed in radio band. In PSRs
B0943$+$10 and B1133+16 for which both $L_x$ and $\hat{P}_3$ are
known, the predicted relationship between observational quantities
holds quite well.

Both the heating and the drifting rate depends on the gap
potential drop. In this paper we made the point that there is
continuum of cases between pure vacuum gap and the space charge
limited flow. The original RS75 model predicts much too high a
subpulse drift rate and an X-ray luminosity. Other available
acceleration models predict too low a luminosity and the
explanation of drifting subpulse phenomenon is generally not clear
(see ZSP05 for more detailed discussion). Approximately, the
bolometric X-ray luminosity for the space charge limited flow
(Arons \& Sharleman 1979) is about $(10^{-4} \div 10^{-5})\dot{E}$
(Harding \& Muslimov 2002), and for the pure vacuum gap (RS75) is
about $(10^{-1} \div 10^{-2})\dot{E}$ (ZSP05), while for the
partially screened gap (GMG03) is $\sim 10^{-3}\dot{E}$ (this
paper). The latter model also predicts right \EB\ plasma drift
rate. Thus, combined radio and X-ray data are consistent only with
the partially screened gap model, which requires very strong
(generally non-dipolar) surface magnetic fields. Observations of
the hot-spot thermal radiation almost always indicate bolometric
polar cap radius much smaller than the canonical value (by an
order of magnitude). Most probably such a significant reduction of
the polar cap size is caused by the flux conservation of the
non-dipolar surface magnetic fields connecting with the open
dipolar magnetic field lines at distances much larger than the
neutron star radius. A small part of this reduction can follow
from the fact that discrete sparks do not heat the entire polar
cap area (say about a half), but uncertainties in determination
the size of the polar cap are likely to account for this effect.

\begin{acknowledgements}
We gratefully acknowledge the support by the WE-Heraeus
foundation. We also acknowledge the support of the Polish State
Committee for scientific research under Grant 1 P03D 029 26.
\end{acknowledgements}

{}

                  \clearpage


\begin{thebibliography}{}
\bibitem{as79} Arons, J., Sharleman,E.T. 1979, ApJ, 231, 854
235,576(CR80)
\bibitem{dr99} Deshpande, A.A., \& Rankin,J.M. 1999, ApJ, 524, 1008 (DR99)
\bibitem{gs00} Gil, J., \& Sendyk, M., 2000, ApJ, 541, 351 (GS00)
\bibitem{gm02} Gil J. \& Melikidze G.I., 2002, ApJ, 577,909
\bibitem{gs03} Gil, J., \& Sendyk, M. 2003, ApJ, 585
\bibitem{gmg03} Gil J., Melikidze G.I., \& Geppert U., 2003, A\&A, 407, 315 (GMG03)
\bibitem{gj69}  Goldreich P., \& Julian H., 1969, ApJ, 157,869
\bibitem{hm02} Harding, A, \& Muslimov,A. 2002, ApJ, 568, 862
\bibitem{kpg06} Kargaltsev,O., Pavlov, G.G., \& Garmire, G.P. 2006, ApJ, 636,
406(KGP06)
\bibitem{l01} Lai D., 2001, Rev. Mod. Phys., 73, 629
\bibitem{n96} Nowakowski,L., 1996, ApJ 457, 868
\bibitem{ml06} Medin, Z., \& Lai D., 2006, astro-ph/0607277
\bibitem{m73} Michel, F.C., 1973, ApJ 180,207
\bibitem{rs75} Ruderman M.A., Sutherland P.G., 1975, ApJ, 196, 51 (RS75)
\bibitem{ug04} Urpin V., \& Gil J., 2004, A\& A, 415, 356
\bibitem{um95} Usov, V.V., \& Melrose, D.B.,1995, Australian J. Phys., 48, 571
\bibitem{um96} Usov, V.V., \& Melrose,D.B., 1996, ApJ 464, 306
\bibitem{wes06} Weltevrede P., Edwards R.I., Stappers B.A., 2005, A\&A, 445, 243 (WES06)
\bibitem{xqz99} Xu, R., Qiao, G.J., \& Zhang, B. 1999, ApJ, 522, L109
\bibitem{zsp05} Zhang B., Sanwal D., \& Pavlov G.G., 2005, ApJ, 624, L109 (ZSP05)
\bibitem{zhm00} Zhang B., Harding A., \& Muslimov A., 2000, ApJ, 531, L135
\bibitem{zh00} Zhang, B., \& Harding, A. 2000, ApJ, 532, 1150
\end{thebibliography}
\end{document}